# Signatures of Star Formation in Brightest Cluster Galaxies

Megan Donahue[a]

[a]Physics and Astronomy Dept., Michigan State University, East Lansing, 48823-2420

**Abstract.** I discuss and review recent studies of the signatures of activity in brightest cluster galaxies. Mid-IR spectra appear to show indications of star formation in a sample of 9 BCGs from de Messieres et al. (2009). Other processes like cosmic ray heating and conduction may play a role. The incidence of emission-line BCGs in X-ray selected clusters is higher than in optically-selected clusters, and higher still in systems known to be cool cores. We report early results of a UV and Hα survey of the BCGs in the REXCESS sample, which reveals that this sample has an interestingly low number of emission-line or UV excess systems.

**PACS:** 95.85Gn, 95.85Hp, 95.85Jq, 95.85Kr, 95.85Mt, 95.85Nv, 97.20.Tr,
97.60.Lf,98.52.Eh,98.54.Ep,98.54.Gr, 98.54.Cm, 98.58.Bz, 98.54.Ca,
98.54.Db,98.58.Jg,,98.62.Ai,98.62.Js,98.62.Mw, 98.62.Qz,98.62.Ve,98.65.Cw,98.65.Hb,

## THE ICM-BCG STAR FORMATION CONNECTION

The state of the hot intracluster medium (ICM) in the core of a cluster of galaxies appears to affect activities, such as star formation and AGN, in the central most massive galaxy, the Brightest Cluster Galaxy (BCG). Decoding the multi-wavelength spectra and photometry of BCGs provides relevant clues about how star formation, accretion of gas, winds, and AGN activity are related in the most massive galaxies in the universe. We could divide BCGs into active and inactive galaxies using any number of criteria from radio luminosities, emission-line sources (including molecular hydrogen), and broad band colors such as UV-optical, or U-B, or optical-IR. They can be very luminous in the radio or the IR. For the purposes of this contribution, we will define an active BCG to have detectable Hα emission lines, with EQW>1 Å.

Clusters appear to be bimodally distributed between low and high core entropy systems [1], and a correlation between the presence of low-entropy X-ray gas in the core of a cluster and the presence of emission line systems [2]. Entropy here is defined to be $K=kT/n_e^{2/3}$, where ne is the electron density in $cm^{-3}$ and kT is the gas temperature in keV

## 1. WHAT IS THE CAUSE OF ACTIVITY IN BCGS?

My bet right now is that the extended line emission is dominated by star formation, but that bet is tempered by knowing the spectra cannot be completely explained by starburst features. Voit & Donahue [3] showed deep optical emission line spectra of

A2597 that was incompatible with shock, and best fit with stellar photoionization plus an extra heating source. AGN photoionization was ruled out by absence of He II recombination. The presence and extent of optical emission lines is correlated with stellar features consistent with recent star formation [4,5,6]. U band polarimetry of A1795 and A2597 excludes scattering from an AGN for these sources [7,8]. The 70-160 micron luminosities of a sample of active BCG are well in excess of stellar continuum, and are correlated with mass cooling rates attenuated to 1-10%, and are strongly correlated with $H_2$ masses [9].

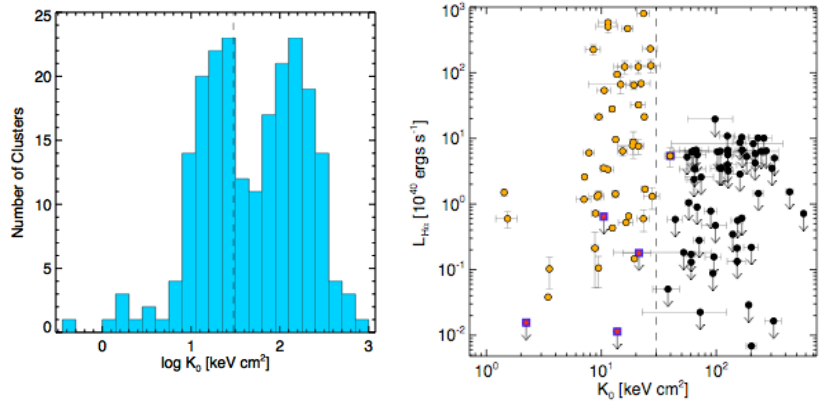

**FIGURE 1.** (a) $K_0$ distribution [1] and (b) $K_0$ and the presence of Hα emission in the BCG [2]. $K_0$ is based on a fit of the radial gas entropy profile, from Chandra archival observations, to a functional form $K(r)=K_0+K_{100}(r/100 \text{ kpc})^\alpha$.

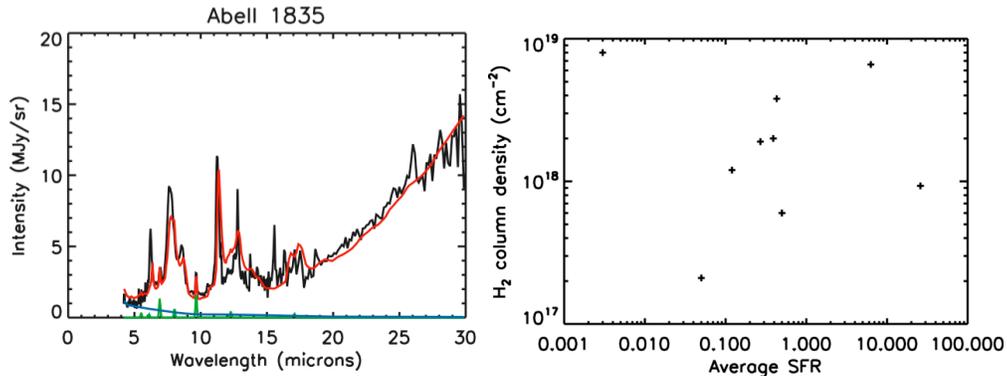

**FIGURE 2.** (a) Spitzer IRS spectrum of the BCG in Abell 1835. Rest-frame wavelength plotted against intensity in relative intensities. (b) $H_2$ column plotted against star formation rate based on mid-IR luminosity. Note that there is no correlation, suggesting separate excitation sources (10% errors in column and $L_{IR}$).

The mid-IR spectra of 9 BCGs [10] (A2597, 2A0335+096, A478, Hydra A, A1068, MS0735, A1795, PKS0745-191, A1835) are well-fit by starburst spectra and the emission lines from molecular hydrogen at ~500-800K (usually requiring 2 temperatures to fit the full line sequence). Figure 2b shows that the $H_2$ luminosities do not correlate with the IR levels (or star formation).

Therefore, it is unlikely that star formation can be left off the list of physical processes occurring in these galaxies. However, the fact that these molecular clouds may be immersed in an environment rather different from that of starburst galaxies or normal galaxies, an environment of hot, high pressure X-ray gas and relativistic particles, means that other processes such as conduction and cosmic ray heating may need to be included as well.

## 2. WHAT IS THE INCIDENCE OF ACTIVITY IN BCGS?

The answer to this question for a representative population of clusters is relevant to testing the timescales for models of activity. If activity is relatively common, it must be rather long-lived. If it is only a feature of local clusters, it may be a recent, late-time feature of BCG evolution. If it is similar in BCGs and ellipticals of similar masses that do not lie in the centers of massive halos, it is simply a product of galaxy-scale environment, not the result of the BCG resting in a special location at the bottom of a $\sim10^{14}$ solar mass gravity well.

It has been known for a very long time that the BCGs in cool core clusters (formerly known as "cooling flow" clusters) frequently exhibit unusual emission line and radio properties. For this contribution, we only summarize Hα results for REXCESS, but we also have UV results in prep [11]. The emission line complex of Ha and [NII] is very easy to detect. Typical sensitivities are EQW>0.5-1.0 Å for Hα, limited by the possibility of a small absorption feature in Ha. [NII] upper limits are even less difficult. So the difficulty in measuring the incidence rate is not in the detection of the emission lines, but in determining the relevant number to use in the denominator, the total number of systems, as we shall see.

The BCGs in B55 sample, the brightest 55 X-ray clusters on the sky as defined by HEAO-1 collimator survey [12] have 21 emission line BCGs and 34 non-emission line BCGs. I culled this information from various sources [13, 14(3C129), 15 (A3532)), and add three non-detections from my own unpublished work (Abell 1650, TriAus, and A4038/Klemola44), for an incidence rate of 38%. [16] record an incidence rate of at least 34% for 38 BCGs with z>0.15, for the BCS/eBCS ROSAT All Sky Survey [17]. [18] report an incidence rate of 30%. The incidence of emission-line BCGs with z=0.2-0.4 in the EMSS is 40-50% [19]. There appears to be very few BCGs with emission lines at z>0.5, but the sample sizes are still quite small, so the statistical significance is limited.

However, our survey of 30 BCGs in the REXCESS sample of representative, unbiased for morphology clusters with z=0.05-0.18, we find an incidence rate of only 2/30, 7%, over 3 sigma different from an expectation of 35%. (We note that the incidence of 20-cm radio sources in or near the REXCESS BCGs is 11/30, 37%., but that only two –the same two- BCGs show significant excess UV emission. [11])

Optically-selected cluster samples tell a different story. [20] report an incidence of only 10% in a "non-cooling flow" sample, an incidence rate no different from the matched (in luminosity) control sample of non-BCG ellipticals. (We note that in this study, the most massive BCGs could not be matched.) The incidence grew to 70-85% for BCGs in their cooling flow sample, while in their optically-selected sample from C4 [21], they find an incidence rate of 15%. This result is very similar to that [22], who report that, in a sample of optically-selected clusters (SDSS C4/DR3 sample), BCGs are more likely to host a radio source than the control sample, are less likely to host an optical AGN than the control sample, and again, an incidence rate of emission lines very similar to the control.

We investigated some of the more obvious possibilities that occurred to us: that the optical samples miss X-ray clusters, that the optical samples have a more difficult time identifying the BCG automatically, that perhaps the unusual nature of the blue-core, optical emission line BCG may exclude it from contention as the BCG in a large survey. What we found was both encouraging and depressing.

1. The overlap between famous X-ray surveys like the B55 and the SDSS cluster coverage from DR5 is tiny. Only 9 z>0.05 B55 clusters are included in the sky coverage of SDSS DR5 (Data Release 7, DR7, will include only a few more.)
2. The C4/DR5 catalogs [23] and the maxBCG/DR5 catalogs [24] find 5 and 4 of the 9 B55 clusters respectively, but together they find 8 of them. They both miss A1689.
3. Whether a cluster is missed or not has no dependence on whether the correct BCG was identified or the incidence of emission lines in the BCG. However, that result has very little statistical power. But if the BCG is identified, the X-ray centroid and the cluster centers are very close in this tiny sample. I would still be worried about color selection and emission line classification effects for a larger sample: the most extreme BCG systems in particular are sometimes classified as "not a galaxy" in the SDSS catalogs.

## CONCLUSIONS

IR (and UV) observations show star formation appears to be a very important process in 1/3-1/2 BCGs in X-ray luminous clusters of galaxies. More correlation studies of large, representative samples of both galaxies selected for activity and control samples, to show trends, as well as detailed deep spectroscopic studies (from the IR through the X-ray) are needed for further progress.

The incidence rate of BCGs with emission lines depends strongly on the parent sample. 35-40% of X-ray bright samples compared to 10-15% of optically selected samples (and those rates are consistent with non-BCG control samples of ellipticals), while ~70% of (strong) cool core systems may host emission-line BCGs.

## ACKNOWLEDGMENTS

I want to acknowledge specific contributions to the projects described in this paper. MSU grad student Seth Bruch carried out SOAR Goodman observations, with assistance from Emily Wang. Emily Wang is completing XMM Optical Monitor UV photometry for REXCESS to satisfy her 2nd year project requirements. Deb Haarsma provided R band photometry for the REXCESS BCGs. Judith Croston and Gabriel Pratt provided derived X-ray properties and additional guidance. University of Virginia graduate student Genevieve de Messieres with her advisor Robert O'Connell provided the IR spectra.. Chris Miller supplied a C4/DR5 catalog shortly after the delivery of his child. Any mistakes or errors in the contribution are mine, but much of the credit belongs to my collaborators.